\documentclass[10pt,sigplan,preprint]{acmart}
\renewcommand\footnotetextcopyrightpermission[1]{} 
\makeatletter
\renewcommand\@formatdoi[1]{\ignorespaces}
\makeatother

\usepackage{booktabs} 
\usepackage{setspace}
\usepackage{url}
\usepackage{amsmath,amssymb}
\usepackage{alltt}
\usepackage{xspace}
\usepackage{epsfig}
\usepackage{graphics}
\usepackage{bbold}
\usepackage{empheq}
\usepackage{xcolor}
\usepackage{multirow}
\definecolor{lightgreen}{HTML}{90EE90}

\usepackage{amsmath,amsfonts,amsbsy}
\usepackage{fancyvrb,keyval,ifthen}
\usepackage{graphicx}

\usepackage{booktabs}
\usepackage{listings}
\usepackage{paralist}
\usepackage{enumitem}
\setlist{nolistsep}
\usepackage{titlesec}
\usepackage{balance}
\usepackage{array}

\usepackage[linesnumbered,ruled,vlined]{algorithm2e}
\SetAlFnt{\footnotesize}

\usepackage{fancyhdr} 
\setcopyright{none}

\makeatletter
\renewcommand{\paragraph}{%
  \@startsection{paragraph}{4}%
  {\z@}{1.5ex \@plus 1ex \@minus .2ex}{-1em}%
  {\normalfont\normalsize\bfseries}%
}
\makeatother

\titlespacing*{\section}
{0pt}{1.5ex}{1.0ex}
\titlespacing*{\subsection}
{0pt}{1.0ex}{1.0ex}

\begin{document}
\title{{\huge A Performance Vocabulary for Affine Loop Transformations}}


\author{Martin Kong}
\orcid{1234-5678-9012}
\affiliation{%
  \institution{Brookhaven National Laboratory}
  \streetaddress{Computational Science Initiative}
  \city{Upton}
  \state{New York}
  \postcode{11973}
}

\author{Louis-No\"{e}l Pouchet}
\affiliation{%
  \institution{Colorado State University}
  \city{Fort Collins}
  \state{Colorado}
  \postcode{80523}
}

\begin{abstract}
Modern polyhedral compilers excel at aggressively optimizing codes with static
control parts, but the state-of-practice to find high-performance polyhedral
transformations especially for different hardware targets still largely
involves auto-tuning.  In this work we propose a novel polyhedral
scheduling technique, with the aim to reduce the need for
auto-tuning while allowing to build customizable and specific transformation
strategies. 
We design constraints and objectives that model several crucial aspects of
performance such as stride optimization or the trade-off between parallelism
and reuse, while taking into account important architectural features of the
target machine.  The developed set of objectives embody a Performance
Vocabulary for loop transformation. The goal is to use this vocabulary,
consisting of performance idioms, to construct transformation recipes
adapted to a number of program classes. 
We evaluate our work using the PolyBench/C benchmark suite and
experimentally validate it against large optimization spaces generated with the
Pluto compiler on a 10-core Intel Core-i9 (Skylake-X).  Our results show that
we can achieve comparable or superior performance to Pluto on the majority of
benchmarks, without implementing tiling in the source code nor using
experimental autotuning.

\end{abstract}

%
%




\renewcommand\acmConference{}{}

\maketitle

\section{Introduction}

Several program optimizations such as loop tiling 
and locality enhancements \cite{uday.pldi.2008,shirako.sc.2014}, automatic
coarse-grained parallelization \cite{sriram.pldi.2007,uday.sc.2012}, 
SIMD-vectorization \cite{kong.pldi.2013,trifunovic.pact.2009,henretty.ics.2013}
and accelerator targeting \cite{grosser.ics.2016,leung.gpgpu.2010,baskaran.ics.2008,ppcg.taco.2013}
have been devised and implemented in research and production compilers using the polyhedral compilation framework \cite{feautrier.ijpp.1991,feautrier.ijpp.1992b}. 
A typical limitation of most, if not all, of these strategies is their limited
ability to adjust the program transformation approach to the specifics of the
input program, and of the target architecture. In practice, typically knobs to
control the internal heuristics are exposed (e.g., the loop fusion scheme, the
tile sizes, etc.), and auto-tuning remains the de-facto approach to achieve
good performance on a variety of targets and programs.

This problem is compounded when considering the variability and trends of upcoming
multi- and many-core processors
where the balance between the hardware parallelism available and the data bandwidth at different memory levels can very significantly differ across architectures, in turn requiring different optimization strategies.
Consequently, performance portability for new hardware targets is often addressed via exploring a search space of candidate transformations, or at worst by a redesign of the program optimization scheme.

The exploration space that a compiler, auto-tuner or expert would have to
consider typically includes loop tiling 
(and the associated tile size selection process),
selecting a good loop fusion/distribution
 structure, the loop permutation order and often some unrolling
factor. 

Polyhedral compilers provide the capability to automatically find an arbitrarily complex loop transformation sequence by reasoning instead on the scheduling of operations, thereby alleviating the phase ordering issue of composing loop transformations.
But severe issues remain, such as
how to achieve seemingly opposing (or at least conflicting) goals:
outer parallelism, inner parallelism for SIMD-vectorization and
locality improvements. 
Each of these topics
has been extensively studied, and several constraints and optimization
criteria have been proposed, e.g. \cite{feautrier.ijpp.1992b,uday.pldi.2008,shirako.sc.2014,uday.sc.2012,kong.pldi.2013,trifunovic.pact.2009,henretty.ics.2013,grosser.ics.2016,ppcg.taco.2013}. Nevertheless, these techniques still employ typically a one-size-fits-all scheduling approach,  suffering from a lack of portability to different architectures.



We propose to tackle this problem by designing
and implementing a {\em performance oriented vocabulary for affine loop transformations}.
In this vocabulary we define several {\em performance idioms},
each of which addresses a specific performance aspect of the program.
The vocabulary can then be used to
\emph{customize the
set of scheduling constraints and objectives to the particularities of the program
and the target architecture}.
In practice, a subset of the performance idioms are embedded into
a Single Integer Linear Program (ILP) \cite{vasilache.07.phd}. 
This type of scheduling technique
starts by constructing the legal space of transformations. In this space,
each point represents a complete set of transformations applied to the code.

Some of the variants in this space could look very similar,
and not differ in anything, 
differ only in the actual coefficients chosen (for instance a skewing
factor) or be drastically different (e.g. two fused serial loops vs. two
distributed parallel loops).  
We use simple metrics and properties of the
code to classify benchmarks into broad categories or program classes, 
and then select the objectives
and constraints that must be embedded into the system.
The ILP is then optimally solved to find in a single step the final program
transformation. This one-shot scheduling approach
\cite{vasilache.07.phd,pouchet.popl.2011} facilitates the encoding of the
properties that we desire to produce on the transformed code.
And essentially, this also allows to order and prioritize the cost
functions according to the set of properties expected on each classification
category.

\begin{figure*}[t]
{\centering
\begin{minipage}{\textwidth}
\begin{minipage}{.37\linewidth}
\begin{tiny}
\begin{lstlisting}
for(t = 0; t < NT; t++){
 for (j = 0; j < NY; j++)
R: ey[0][j] = data[t];
 for (i = 1; i < NX; i++)
  for (j = 0; j < NY; j++)
S: ey[i][j] = ey[i][j] - 0.5*(hz[i][j]-hz[i-1][j]);
 for (i = 0; i < NX; i++)
  for (j = 1; j < NY; j++)
T: ex[i][j] = ex[i][j] - 0.5*(hz[i][j]-hz[i][j-1]);
 for (i = 0; i < NX-1; i++)
  for (j = 0; j < NY-1; j++)
U: hz[i][j] = hz[i][j] - 0.7 * (ex[i][j+1] 
     - ex[i][j] + ey[i+1][j] - ey[i][j]);
}
\end{lstlisting}
\end{tiny}
\end{minipage}
\hspace{2pt}
\begin{minipage}{.34\linewidth}
\includegraphics[width=0.98\textwidth,height=1.35in]{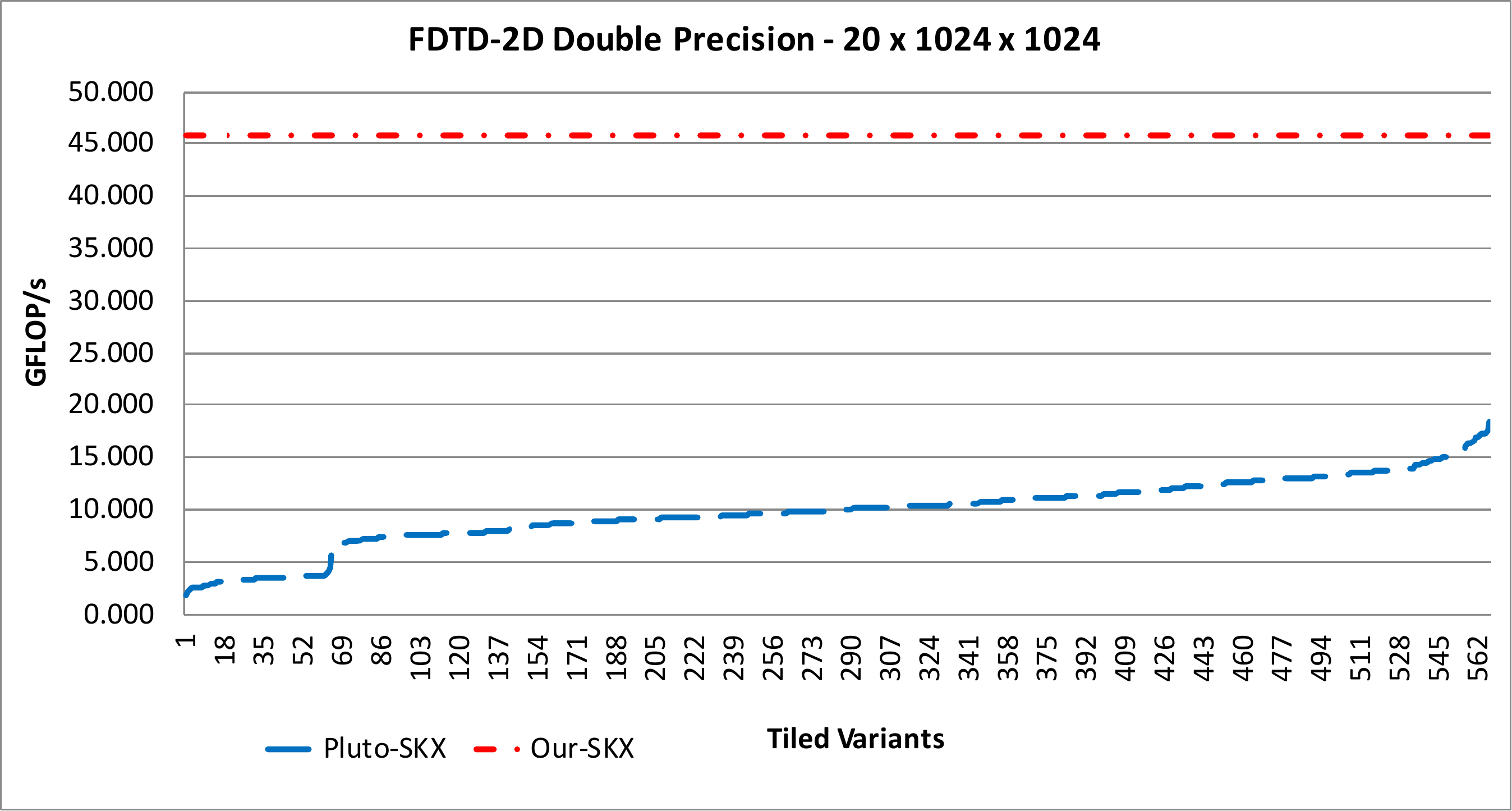}
\end{minipage}
\hfill
\begin{minipage}{.27\linewidth}
{%
\begin{tiny}
\begin{tabular}{|c|c|c|}
\hline
Metric & our & pluto  \\
\hline
Load to store ratio   & 3.1368 & 3.6534 \\
Retired mem.loads   & 1.14E+7 & 5.65E+7 \\
Retired mem.stores   & 3.69E+6 & 1.55E+7 \\
L2 LINES IN         & 9.86E+5 & 7.08E+5 \\
L2 to L1 load  [GBytes]     & 0.4298 & 0.1745 \\
L1 to L2 evict  [GBytes]    & 0.1327 & 0.0866 \\
L3 to L2 load  [GBytes]     & 0.0631 & 0.0453 \\
L2 to L3 evict  [GBytes]    & 0.0112 & 0.0745 \\
System to L3  [GBytes]     & 0.0585 & 0.0371 \\
L3 to system  [GBytes]     & 0.0026 & 0.0030 \\
L2 request rate     & 0.4303 & 0.1045 \\
L2 miss ratio      & 0.0806 & 0.2378 \\
Vectorization ratio   & 94.0256 & 0.0022 \\
Operational intensity         & 0.3782 & 0.4626 \\
Avg stall duration [cycles]   & 25.2594 & 15.8386 \\
\hline
\end{tabular}

\end{tiny}
}
\end{minipage}
{%
\begin{footnotesize}
\caption{\label{fig:fdtd2d-motiv} FDTD-2D kernel: [left] source code; [center]
performance distribution of tiled variants on Core-i9 (SKX); [right] hardware
counters comparison of best Pluto tiled variant and our apporach, on a 10-core i9 (SKX)}
\end{footnotesize}
}
\end{minipage}
}
\end{figure*}

In summary, we make the following contributions:
i) we model in a single ILP numerous novel constraints and objectives
for simultaneously 
extracting outer and inner (SIMD-vector) parallelism, trading 
parallelism for reuse, reference-level stride optimization 
and a fusion/distribution approach;
ii) we develop a \emph{customized}
scheduling approach that modifies the set of objectives and constraints used in the ILP as a function of the specifics of the benchmark and the target architecture, and provide evidence that classifying the type of kernels
according to simple metrics extracted from affine 
programs can be utilized to judiciously enable or disable constraints
and objectives of the ILP;
iii) we propose a novel unroll-and-jam cost model that maximizes
cache-line usage and that accounts for outer parallelism and inner 
data reuse;
%
%
iv) we perform an extensive experimental 
evaluation on a modern multi-core processor,
and demonstrate that we can achieve 
comparable or superior performance to the Pluto compiler
for most benchmarks evaluated,
all without resorting to source code tiling nor autotuning.

The rest of this paper is organized as follows. The next section 
discusses the motivation of our work. Section \ref{sec:background}
presents the background concepts of the polyhedral model. Section 
\ref{sec:transfos} describes the ILP formulation
implemented in our transformation engine. 
Section \ref{sec:results} evaluates the efficiency
of our transformations. Finally, Section \ref{sec:related}
and Section \ref{sec:conclusion} discuss the related work and the
conclusions of this paper, respectively.

\label{sec:intro}

\section{Motivation}

Modern multi and many-core processors such as IBM Power9, Intel KNLs and
Skylakes, exhibit sufficiently distinct architectural features from their
earlier multi-core predecessors. Some of the most strikingly different features
are much larger L2-caches, which in Skylake's becomes a 1MB private cache per
core. Substantially larger L3 caches are also on the rise. 
Another important feature to consider is
hardware parallelism, i.e. the number of cores and the width of the vector
registers.


Figure \ref{fig:fdtd2d-motiv} shows the source code of the FDTD-2D kernel from
the PolyBench/C benchmark suite. The plot on this figure represents the
performance distribution of 567 different tile configurations by using the
Pluto polyhedral compiler, and applying the wave-front parallelization
strategy. These variants were ran on 
a 10-core Intel Core-i9 7900X.
This performance looks far from impressive. 
As a reference, a $1024^3$ DGEMM using LAPACK on this machine achieves
0.6 TF/s.

To achieve this performance, we have designed and developed a novel set of
performance constraints and objectives on a polyhedral compiler. Each of the
implemented objectives fulfills a specific role and purpose in the overall
approach. Collectively, they comprise a {\em Performance Vocabulary for Affine
Transformations}, and are meant to be used for a wide variety of program
classes and architectures. Although this set is not comprenhensive yet, they
can be seemlessly combined and used to build more powerful transformations.
These are embedded into a single integer linear program (ILP) and allow to
model several relevant features such as: a) inner-vector parallelism, b)
inner-loop stride access, c) selection of outer parallel loop dimension and
trade-off between inner data reuse, d) skewing minimization for stencil
computations, e) automatic loop fusion/distribution selection, f) separation of
independent statements and non-flow related statements, g) true dependence
distance minimization.  Unlike previous works, these objectives and constraints
are not applied indiscriminately. Instead, we carefully craft and apply
``transformation recipes'' depending on a number of metrics extracted from the
input program as well as the target architecture. To give a simple example,
objective (c) is not essential for performance on a many-core processor such as
Intel's KNL or IBM's Power9, but it is fundamental to achieve a high-fraction
of the machine's effective peak performance on the Skylake processor. More
importantly, our approach does not require extensive and costly empirical space
exploration or autotuning in order to achieve this performance, nor does it use
the loop tiling (blocking)  transformation to achieve high-locality.

The hardware counters in (Table in Fig. \ref{fig:fdtd2d-motiv}) 
compare our transformed program
to the best performing pluto variant in the space explored.  
Some differences are rather surprising. For instance, the best pluto
variant incurs in 5x more memory loads and 4.2x more stores than our transformed
code. 
Note in contrast to Pluto we do not use loop tiling, which often result in code more difficult for the back-end compiler to optimize, losing vectorization potential.
Nonetheless, the ratio between incoming
and evicted data for our optimized code is 3, 5 and 22, for the L1, L2
and L3 caches. This is a much higher ratio than Pluto's variant, which
achieves ratios of 2, 0.6 and 12. This means that our transformation strategy,
despite inducing a higher total volume, makes better use of the data brought in.
Next, despite a 4x higher L2 request rate, we achieve 3x fewer misses on the L2 
cache. 




\label{sec:motivation}

\section{Background\label{sec:background}}
Here we recap background concepts of the polyhedral
model that we use in Section \ref{sec:transfos}, namely iteration
domains, access functions, dependence polyhedra, program schedules
and the notation used to model our constraints and objectives.
\textbf{Iteration domains:}
In the polyhedral model, each syntactic statement $S$ is associated
to a set $\mathcal{D}^S \in \mathbb{Z}^{+}$ 
comprised by the
dynamic instances of the statement.
Then, via \textbf{Access functions},
each iteration domain is mapped 
to the (multi-dimensional) data-space of arrays referenced in them.
Each access function $F^A$ is represented as a $M_A \times N_S$ 2D array,
where $M_A$ is the number of space dimensions of the array and $N_S$
is the number of loops surrounding statement S.
In our FDTD-2D example, the first read reference of array $ex$ in statement $U$ is
represented as 

{
{%
\begin{footnotesize}
{%
\noindent
\begin{equation*}
F^{ex}(\vec{x}) = \left[ 
\begin{matrix}
0 & 1 & 0 & 0 & 0 & 0 & 0\\
0 & 0 & 1 & 0 & 0 & 1 & 0\\
\end{matrix}
\right]
(t~i~j~NT~NX~NY~1)^T = 
\left[
\begin{matrix}
i \\ 
j + 1\\
\end{matrix}
\right]
\end{equation*}
}
\end{footnotesize}
}
}

\textbf{Dependence polyhedra}
embody the semantic orderings of a program.
Every program dependence in a SCoP is represented by one or more
dependence polyhedra $\mathcal{D}^{R,S}$. These polyhedra define
the ordering among points $\vec{x}^{R}$ and $\vec{y}^{S}$
from the iteration domains $\mathcal{D}^{R}$
and $\mathcal{D}^{S}$, respectively.
Lastly, \textbf{program transformations}
are represented in the polyhedral model by
scattering functions. These functions assign to each point of the iteration
domain
$\mathcal{D}^{S}$ a timestamp which determines its execution order.
Timestamps can be multi-dimensional and are implemented 
as matrices. The number of rows of the matrix must be defined 
a priori. For this work we use the {\bf 2d+1} representation, which
assumes that the scattering matrix will consist of {\bf 2d+1} rows
and {\bf d+1} columns,
where $d$ is the maximum loop depth.
The advantage of this representation is its natural mapping to the
syntactic nesting structure. Assuming the schedule ie zero-offset, 
the even rows are known as scalar dimensions
and represent the nestingness and order of statements and loops,
whereas the odd dimensions represent linear transformations (e.g. skewing
or loop permutation) applied to the input code.
Below we show the scheduling matrices for statements R and S
of our FDTD-2D example.

{%
\noindent
{%
\begin{scriptsize}
\begin{equation*}
\Theta^R = 
\begin{pmatrix}
0 & 0 & 0\\
1 & 0 & 0\\
0 & 0 & 0\\
0 & 1 & 0\\
0 & 0 & 0\\
0 & 0 & 0\\
0 & 0 & 0 
\end{pmatrix}
\begin{pmatrix}
t \\
j \\
1 
\end{pmatrix}
=
\begin{pmatrix}
0 \\
t \\
0 \\
j \\
0 \\
0 \\
0
\end{pmatrix}^T,
\Theta^S = 
\begin{pmatrix}
0 & 0 & 0 & 0\\
1 & 0 & 0 & 0\\
0 & 0 & 0 & 1\\
0 & 1 & 0 & 0\\
0 & 0 & 0 & 0\\
0 & 0 & 1 & 0\\
0 & 0 & 0 & 0 
\end{pmatrix}
\begin{pmatrix}
t \\
i \\
j \\
1 
\end{pmatrix}
=
\begin{pmatrix}
0 \\
t \\
1 \\
i \\
0 \\
j \\
0
\end{pmatrix}^T
\end{equation*}
\end{scriptsize}
}
}

\paragraph*{Notation}
In the rest of this paper, $R$ and $S$ denote arbitrary statements,
$N_S$ is the number of statements in the SCoP, $F, F^A$ are generic
access functions, d and dim($\theta$) are the No. of rows in $\theta$,
{\bf dim} is a function that returns the number of loops surrounding
S and the number of space dimensions in F, $I^S$ is the identity 
schedule for statement S. A linear dimension of $\theta$ is any
row-i s.t. $i \mod 2 = 1$, while for a scalar dimension this is
$i \mod 2 = 0$. Lastly, $\beta^S$ is the portion of the schedule
corresponding to the last column of $\theta^S$.


\section{Transformations\label{sec:transfos}}

Prior approaches for polyhedral scheduling typically aims for the best
one-size-fits-all formulation: irrespective of the input program, the same
scheduling problem is solved \cite{uday.pldi.2008,kong.pldi.2013}. Our work
takes a significantly different approach: instead of attempting to find the
best unique set of constraints/objectives to optimize for, we instead develop a
catalogue of individual smaller scheduling formulations, each targetting a
specific performance objective. We then select, as a function of the input
program, which of these shall be used, \emph{and in which order of priority}
they shall be solved. 

We consider the following optimization objectives: outer parallelism, inner
(SIMD) parallelism, parallel dimension selection and inner reuse tradeoff,
stride access minimization and stencil skewing parallelization, independent
statement separation and dependence guided fusion. We next briefly recap the
single-ILP legal space construction we build our work on
\cite{pouchet.popl.2011,kong.pldi.2013}, before detailing the formulation of
these individual objectives.

In this work we seek to avoid the hassle of experimental exploration very often
required in HPC and in polyhedral compilation frameworks. In particular, as we
will see in Sec. \ref{sec:results}, we intentionally compare a single non-tiled
variant to an autotuned space which includes 3 distinct fusion and heuristics
and several tile sizes for each of these (Note: a fully integrated approach to
combine our current goals plus the tiling constraints are left for future
work). The nature of our main goals, i.e. achieving outer and innermost
parallelism in addition to good locality usually embody opposing and
conflicting goals.  A scheduling approach such as Feautrier's
\cite{feautrier.ijpp.1992a,feautrier.ijpp.1992b} which maximizes the dependence
satisfaction at the outermost scheduling dimensions directly favor schedules
which could potentially have multiple degrees of inner parallelism.
Nevertheless, it also impacts the generated code by very often resulting in
skewing transformations.  On the other hand, the state-of-the-art Pluto
polyhedral compiler and its tiling hyperplane method in order to find maximal
tiling hyperplanes, push the dependence satisfaction into the innermost
dimensions. This commonly results in tiled codes which are inherently serial.
Approaches such as \cite{kong.pldi.2013} take a decoupled approach which
combines tiling and exposing coarse grained parallelism proceeded by a second
stage which restructures the full and partial tiles of the parametrically tiled
code. As we will see, the same set of constraints and objectives are not always
necessary to achieve good performance. A good and simple metric and few
decisions that select that nature of the constraints to be embedded should
suffice to explore the legal space which in itself is an optimization / tuning
space.

\subsection{Building the legal space}
\label{sec:transfos:legal}
We first 
construct the {\bf convex set of semantics preserving transformations} \cite{vasilache.07.phd,pouchet.popl.2011}.  The
ILP system built for this purpose integrates legal constraints for all
dependence polyhedra of the original program. Given a point $\langle x_R, y_S
\rangle \in \mathcal{D}_{R,S}$, the semantics of the original program state
that the statement instance $x_R$ executes before $y_S$. Thus, in the
transformed program, the schedule applied to both of these statements,
$\Theta^R$ and $\Theta^S$ must preserve the ordering condition $\Theta^R(x_R) \prec
\Theta^S(x_S)$, which states that the timestamps assigned to them must maintain
their relative order. As we use multi-dimensional schedules of depth 2d+1, this
means that the difference between the two schedules will be a lexicographic
positive vector: $\Theta^S(y_S) - \Theta^R(x_R) \succ
(\delta_1,\delta_2,..,\delta_{2d+1})$.  In practice, we define 2d+1 $\delta$
variables for each dependence polyhedron, and bound each
$\delta^{\mathcal{D}_{R,S}}_i, ~~ i \in \{1,..,2d+1\}$ to the set \{0,1\},
making it a boolean variable. This way, a dependence is satisfied when any of
the $\delta_i$ variables become positive.  These constraints are formalized as:
\noindent
\begin{equation}
{\footnotesize
\label{eq:legal-space}
\boxed{
\begin{split}
\forall \mathcal{D}_{R,S}, \forall l \in 0..d-1, \delta^{\mathcal{D}_{R,S}}_{l} \in \{0,1\},
\forall \mathcal{D}_{R,S}, \sum^{d-1}_{l=0} \delta^{\mathcal{D}_{R,S}}_{l} = 1 
\\
\forall \mathcal{D}_{R,S}, \forall l \in \{0,..,d-1\}, \forall \langle\vec{x}_{R},\vec{y}_{S}\rangle \in \mathcal{D}_{R,S}: \\
\Theta^S_l({\vec{x}_{S}}) - \Theta^R_l({\vec{y}_{R}}) \ge -  
\sum^{l-1}_{c=0} \delta^{\mathcal{D}_{R,S}}_{c} \cdot ( K \cdot \vec{n} + K ) + \delta^{\mathcal{D}_{R,S}}_{l}
\end{split}
}
}
\end{equation}

In order to not over-constrain the legal space, as soon as a dependence is
satisfied at some schedule level $l$, we nullify the constraints for all
subsequent dimensions $c > l$. This is achieved in Equation \ref{eq:legal-space}
by the term $K \cdot \vec{n} + K, K \in \mathbb{Z}$.
Coefficients $\theta^S_{i,j}$ are bounded, for a sufficiently large K
\cite{pouchet.popl.2011}, this expression upper bounds the time elapsed
between any two statement instances. Thus, when $\delta_l = 1$, for some $l$,
the sum of the right-hand side of the inequality becomes in practice $-\infty$.
On the contrary, all levels $c < l$ would have $\delta_c = 0$, so the sum
becomes zero, making the constraint valid.
Lastly, we apply the affine form
of the Farkas Lemma \cite{feautrier.ijpp.1992b} to compute all the
non-negative functions over the candidate system. Additional constraints and objectives can then be embedded to then find good program schedules \cite{feautrier.ijpp.1992b,kong.pldi.2013}.

\subsection{Outer Parallelism (OP)}
\label{sec:transfos:outerpar}
As we will see, in some scenarios it's obvious 
that extracting OP is possible (e.g. DGEMM),
while in others (e.g. LU), is necessary to use the
second outermost loop. In both cases
we use the $\delta$ variables embedded into the
convex space, and minimize their sum at a predefined
level selected in the following fashion:
\noindent
\begin{equation}
{\footnotesize
\label{eq:outerpar-lin1}
\min \sum_{\mathcal{D}_{R,S}} \delta^{\mathcal{D}_{R,S}}_{p}: if ~N_{SCC} \geq N_{self.dep}~p=1~ else~p=3
}
\end{equation}
Dimension-p is parallel when this sum is 0. However,
a non-zero sum doesn't necessarily preclude extracting parallelism.

\subsection{Stride Optimization (SO)}
\label{sec:transfos:strideopt}

To maximize the vectorization potential and program locality
we 
drive the selection of transformations
to minimize high-strides of innermost loops. 
The program objective adds the costs of all individual
statements that have dimensionality 2 or higher, 
while the statement stride cost
is built from a weighted sum of its references.
Unlike previous works \cite{vasilache.impact.2012,kong.pldi.2013,verdoolaege.impact.2018} 
we don't aim to maximize stride-1/0
references. Rather, we aim to minimize high-stride
penalties as a whole by using
higher penalties for stride-0 references than for stride-1, 
and doubling the penalty for write references. 
The reason for this is that if the iterator
originally bound to the fastest varying dimension (FVD) of the array
does not appear in $\theta^S_{d-2}$, 
then some other iterator will take its place,
thereby inducing a high-stride if the new iterator appears
in any non-FVD.
This is possible when, for instance, a loop permutation
is applied.

To minimize high-strides we compute a set of
weights for each loop iterator and reference. 
These weights represent the stride cost when
the loop associated to a designated iterator is 
moved into the innermost loop dimension: 
\noindent
\begin{equation*}
{\footnotesize
\label{eq:iterator-stride}
\boxed{
\begin{split}
\forall S, \forall it \in 0..dim(S)-1: W(S,it) = \sum_{F \in S} W(F, it) ~.~ P(F)
\\
W(F,it) = 
\left\{
\begin{matrix}
  1 : & {\bf it} ~~ \in ~~ FVD(F); \\
  3 : & {\bf it} \not\in F; \\
  10 : & {\bf it} \in F \wedge {\bf it} \not\in FVD(F)
\end{matrix}
\right\}
, P(F) = 
\left\{
\begin{matrix}
1 : F~is~read\\
2 : F~is~write
\end{matrix}
\right\}
\\
\forall S: cost(S) = \sum_{k=0}^{dim(S)-1} \theta^S_{d-2,k}~. ~W(S,k)
\end{split}
}
}
\end{equation*}

Next, we create two new variables per statement, 
one that minimizes the sum of $\theta$ coefficients, and another
to aggregate the stride cost.
These variables are
placed in the leading position of our system to give them the highest
possible priority. Thus, the objective for each statement becomes:
\noindent
\begin{equation}
\label{eq:program-stride}
\min \left\{ \sum^{dim(S)-1}_{k=0} \theta^S_{d-2,k},~~ \sum_{S} cost(S) \right\}
\end{equation}

\subsection{Inner Parallelism (IP)}
\label{sec:transfos:innerpar}

Inner parallelism is modeled in a manner analogous to outer 
parallelism, except for the schedule dimension(s) corresponding 
to inner loops. The number of loop-carried dependences at the 
inner-most loop level shall be 0 if possible to achieve a parallel loop, 
which means they shall be carried at outer levels \cite{feautrier.ijpp.1992a}. We only seek inner-parallelism when the 
loop depth is 3 or more, as 1D and 2D parallel loop nests are already covered by the outer parallelism case. 
Note an outer-parallel loop can always be sunk inner-most, possibly after using strip-mining \cite{uday.pldi.2008}.

\noindent
\begin{equation}
\begin{split}
\label{eq:innerpar}
\min \sum_{\delta^{\mathcal{D}_{R,S}}} \delta^{\mathcal{D}_{R,S}}_{2d-1} 
\end{split}
\end{equation}

\subsection{Outer Parallelism and Inner Reuse (OPIR)}
\label{sec:transfos:outerpar-inner-reuse}

To model the performance trade-off
between parallelism and reuse we focus on the
schedule dimensions of each statement $S$ that model loops (i.e., the odd rows),
and bind these to
the spatial dimensions of each array reference $F \in S$.
In practice, this novel formulation usually boils
down to finding the best loop permutation that 
optimizes reuse while simultaneously
selecting the best schedule dimension for outermost
parallelism. We construct a set
of auxiliary functions for each non-scalar reference: 
$M^F_{1 \times dim(S)}$,
$G^F_{d \times dim(S)}$ and $R^F_{1 \times dim(S)}$.

\noindent
\begin{equation*}
{\footnotesize
\label{eq:reference-mat-MGR}
\boxed{
\begin{split}
\forall S, \forall F \in S, k \in 0..dim(S)-1:
M^{F}_k = \sum_{i = 0}^{dim(F)-1} | F_{i,k} |
\\
\forall S, \forall F \in S, \forall i \in 0..\min\{dim(F),dim(S)\}-1, \forall j \in 0..dim(S)-1: 
\\
G(F,M^F)_{i,j} =
\left\{
\begin{matrix}
  M^F_{j} : & M^F_{j} > 0 \wedge F_{i,j} \neq 0 \\
  - 1     : & M^F_{j} > 0 \wedge F_{i,j} = 0 \\ 
    0     : & otherwise
\end{matrix}
\right\}
\\
 \forall F, \forall j \in 0..dim(S)-1: 
R(M^F)_{j} =
\left\{
\begin{matrix}
  \lfloor d/2 \rfloor - j: & M^F_{j} > 0\\
  0 : & otherwise
\end{matrix}
\right\}
\end{split}
}
}
\end{equation*}

The above matrices are constructed for each array
reference $F \in S$ in the SCoP.
Their roles are the following:
matrix $M^F$ computes 
weights of the iterators on reference $F$, 
matrix
$G^F$ represents the parallelism component and its
relation with $F$'s space dimensions, 
and $R^F$
embodies the potential reuse in an array w.r.t.
the schedule of statement S:




Bounds on the schedule 
$\theta^S$ are computed by taking the identity schedule ($I^S$)
as the reference,
and schedule rows of
non-full dimensional statements are explicitly padded. 
Not doing
so allows the scheduler to pad any of the outer rows, disrupting our objective.
These constraints are shown below:
\noindent
\begin{equation*}
{\footnotesize
\label{eq:parallelism-vs-reuse-constraints}
\boxed{
\begin{split}
\forall S, k \in 0..d-1, k~~odd: 
\left( \sum_{j = 0}^{dim(S)-1}  I^{S}_{2k+1,j} \right) -
\left( \sum_{j = 0}^{dim(S)-1} \theta^S_{2k+1,j} \right) \geq 0 
\\
\forall S, dim(S) < d, \forall i \in 0..d-dim(S)-1, \forall j \in 0..dim(S)-1:\\
\theta^S_{2i+1,j} = \theta^S_{2(d-dim(S))+1,j}
\end{split}
}
}
\end{equation*}

\paragraph{{\bf Q: between (P)arallelism and (R)euse}}
We now have all the necessary ingredients to complete our
formulation for trading parallelism for inner data reuse.
For each (non scalar) array reference $F$ (appearing 
in some statement S) and linear scheduling dimension $i$, 
we create a pair of variables,
$Q^{+F}_i$ and $Q^{-F}_i$. These variables relate
three aspects of our formulation:
parallelism, schedule-to-data-space
mapping and the reuse potential. 
The first component is represented in our formulation by
the $1 - \delta^{\mathcal{D}^{S,S}}_{i}$ term, which equals 0
when a dependence is satisfied at schedule
dimension $i$. The second component makes use
of the auxiliary matrix $G(F,M^F)_{i,j}$. Its role
is two-fold: i) mapping
outer loop dimensions of $\theta^S$ to spatial dimensions 
of array $F$; ii)
reward or penalize transformations depending 
on the appearance of specific iterators in $F$
and that are used (or not) in $\theta^S$.
The third component further rewards transformations
where the loop iterators that appear in $\theta^S$
are not used
in reference $F$, with higher rewards given to
outer loop dimensions (see definition of
array $R^F_j$ above). 
This essentially is the reuse component
of the formulation. 
Lastly, the $padding$ variable is computed per statement,
and is only required when statement $S$ is not full-dimensional
(i.e. $dim(\theta^S) < d$).
\noindent
\begin{equation*}
{\footnotesize
\label{eq:parallelism-vs-reuse-Q}
\boxed{
\begin{split}
 & \forall S, F \in S, C \equiv min\{dim(S),dim(F)\}-1, i \in 0..C:
 \\
Q^{F}_i \leq ~
& \left( 1 - \delta^{\mathcal{D}^{S,S}}_{2i+1} \right) ~ + ~
\left( \sum_{j = 0}^{dim(S)-1} 
 G(F,M^F)_{i,j} ~. ~\theta^S_{2i+1,j} \right) ~ + \\
& \left( \sum_{j = 0}^{dim(S)-1} \sum_{k = i + 1}^{C} R(M^F)_j ~ . ~ \theta^S_{2(padding+k)+1,j} \right)
\end{split}
}
}
\end{equation*}

Here we give a very brief example.
Suppose we have the statement
\texttt{C[i][j] += A[i][k] * B[k][j]} from DGEMM,
and that row-5 of $\theta^S$ has been
set to [0,1,0] by the previous objectives.
The M-matrices for C, A and B would be
[1,1,0], [1,0,1] and [0,1,1], respectively.
Similarly, the G-matrices  would be [1,-1, 0;-1,1,0], [1,0,-1;-1,0,1] and
[0,-1,1;0,1,-1]. The reader can notice
that each row of G$^F$ can also
be obtained by using M$^F$ as a mask
over each row of F: if the corresponding
values are positive, the value of M is
used; if M is positive and F is 0, then -1 
is set in G; for any other case, then 
entry of G becomes 0. The R-matrices
simply represent the weight of each 
schedule dimension. Hence we have
$R^C=[3,2,0]$, $R^A=[3,0,1]$ and 
$R^B=[0,2,1]$. All these factors are used
in the above constraint. The reader
can observe that when $\delta_1=0$
(outermost parallel loop), the first
two rows of $\theta^S$ become [1,0,0;0,0,1].
On the other hand, when $\delta_1=1$,
$\theta^S$ becomes [0,0,1;1,0,0].
Between these two options, the latter
is the one that maximizes the
overall cost of the statement (See next
set of constraints). This pushes up
the values of its $Q^F_i$, and in turn,
make $\delta_1=1$.

When a statement carries no self-dependence, the $1 - \delta$
term is removed,
and the constraints 
attempt to find the best loop order that
minimizes the number of high-strides in statement S.
Next, the cost functions of parallelism-vs-reuse
are aggregated per statement $S$. Each $Q^{-S}$ and
$Q^{+S}$ are bounded by a compile-time
constant that varies per statement ($UB^S$):
\noindent
\begin{equation*}
{\footnotesize
\label{eq:parallelism-reuse-objective}
\boxed{
\begin{split}
\forall S: 
UB^S = \sum_{F} \sum_{i = 0}^{C} (2 + \lfloor {d/2} \rfloor - i)
\wedge
Q^{+S} = \sum_{F} \sum_{i} Q^{F}_{i} \wedge \\
0 \leq Q^{-S}, Q^{+S} \leq UB^S
\wedge
Q^{prog} = \sum_{S} Q^{-S} 
\end{split}
}
}
\end{equation*}
Finally, the overall
program is optimized by minimizing the sum
of $Q^{-S}$ variables into $Q^{prog}$ and inserting 
it into the leading position of the current system. 
This, in turn, maximizes the
values of the $Q^{+S}$ variables in our formulation:
\noindent
\begin{equation}
\min \left\{ Q^{prog}\right\}
\end{equation}

\subsection{Dependence Guided Fusion (DGF)}
\label{sec:dep-guided-fusion}

To achieve even better locality, we enhance our formulation 
with constraints and objectives for fusion {\em driven
only by inter-statement flow-dependences} (ISFD).
WAR and WAW dependences are not considered to alleviate
pressure on the back-end register-scheduler (In the next
section we actually increase the lexicographic distance
between statements), while we consider RAR dependences
to be unprofitable in our scheme unless full fusion
is achieved. We thus exclude it as well.



The equations below are embedded in the ILP for each 
candidate dependence in the program.  
We skip all other types
since these have little
influence in the producer-to-consumer distance between two statements in a
dependence relation. Before constructing the below 
constraints, we build the strongly connected components
(SCC) graph, and query it to decide if two statements
are separable
This step is independent of the actual
fusion structure of the input program. 
The focus of this objective is to capture
inter-statement reuse, since our OPIR and SO objectives
already account for intra-statement reuse.
Moreover, we have observed that
intra-SCC + inter-statement reuse is not beneficial
since the statements will already share the outermost loop dimension.
In practice, this is not a problem unless using extremely
large problem sizes (e.g. 32KB or 64KB on a single array
dimension).

\noindent
\begin{equation*}
{\footnotesize
\label{eq:dependence-fusion}
\boxed{
\begin{split}
\forall \mathcal{D}_{R,S}, R \neq S, FLOW(\mathcal{D}_{R,S}) 
\equiv True, SCC(R) \neq SCC(S), 
\\
dim(R,S) \equiv min\{dim(R),dim(S)\} - 1
:
\\
\Delta^{\mathcal{D}_{R,S}} = \sum_{i = 0}^{dim(R,S)} 
(\beta^S_{2i} - \beta^R_{2i} ) ~ . ~ W_i
\\
\forall \Delta^{\mathcal{D}_{R,S}}: 0 \leq \Delta^{\mathcal{D}_{R,S}} \leq No.stmts ~ . ~ 2^{\lceil d/2 \rceil + 1}
\\
\forall i \in 0..dim(R,S): W_i = 2^{\lceil d/2 \rceil  - i - 1}
\end{split}
}
}
\end{equation*}

We assign to each component of $\vec{W}$ 
a distinct power-of-2 weight.
A similar scheme was first used by Vasilache
\cite{vasilache.07.phd}.
If an  ISFD
involves statement R and S, then the above equation assigns a higher
cost to the outer scalar dimensions. This cost function is minimized by
bringing closer the scalar dimensions of the respective statements R and S.
Clearly, when two statements share the same scalar coefficient at some level k,
the term associated to this level becomes zero. As a final touch,
when the array reference of the flow dependence {\bf is }
simultaneously a write-reference of the dependence target, then we double
the cost of every term in $\vec{W}$. Finally, we minimize the sum of $\Delta$ variables and insert the sum variable in the leading position of the
current system:
\noindent
\begin{equation}
{\footnotesize
\label{eq:obj:dgf}
\min  \{\sum_{\mathcal{D}_{R,S}} \Delta^{\mathcal{D}_{R,S}}\}
}
\end{equation}

\subsection{Separation of Independent Statements (SIS)}
\label{sec:statement-separation}

A weakness of dependence distance optimization schemes based on the
minimization of the maximal reuse distance \cite{uday.pldi.2008}
is the inability to distinguish between
dependences that affect performance and those that do not. Fusion
of unrelated statements can be very detrimental to performance as
these can essentially flush the cache. Thus, we introduce the concept
of {\bf independence distance}. The intuition is the following:
dependence distance minimization attempts to bring closer all statement
instances in a dependence relation, zipping them together from the 
outermost scalar dimension to the innermost. Now, as a given 
statement S can be the sink of several dependences (RAW, WAR, WAW and RAR),
the statement that ends up closer to S could be practically 
anyone which is a source of a dependence. This order will
ultimately be determined by several factors including the order
in which the dependences are processed. 

\noindent
\begin{equation*}
{\footnotesize
\label{eq:independent-map}
\boxed{
\begin{split}
I_{R,S} =
\left\{
\begin{matrix}
True : & (FLOW(\mathcal{D}_{R,S}) \equiv True \wedge 
\\
       & R \neq S, SCC(R) \neq SCC(S)) \vee 
(\neg\exists \mathcal{D}_{R,S})
\\
False : & otherwise
\end{matrix}
\right\}
\\
\forall S: 0 \leq \beta^S_0 \leq N_S
\wedge \sum_{S} \beta^S_0 \leq N_S \times (N_S + 1) / 2
\\
\forall R,S, I_{R,S} \equiv True, R < S:
0 \leq \nabla^{-}_{R,S}, \nabla^{+}_{R,S} \leq S - R
\\
\nabla^{-}_{R,S} + \nabla^{+}_{R,S} = S - R
\wedge
\nabla^{+}_{R,S} = \beta^S_0 - \beta^R_0
\\
\nabla_{sum} = \sum \nabla^{-}_{R,S}
\wedge
0 \leq \nabla_{sum} \leq N_S
\end{split}
}
}
\end{equation*}

To maximize the {\bf good} fusion potential in the program
we increase the distance between independent statements.
Coupling {\bf SIS} with the {\bf DGF} objective has proven to 
produce robust schedules with good locality in our evaluation.
We follow three criteria to decide when to introduce our separation constraints
and objectives:
i) $\exists \mathcal{D}_{R,S}$ and it is not a flow dependence, or 
$\neg\exists\mathcal{D}_{R,S}$;
ii) $R \neq S$; and
iii) both R and S do not belong to the same SCC. (Note: in the equations
below, R and S are both statements as well as natural numbers 
with values equal to their program order location):


We build a set $\mathcal{I}$ with all the pairs of statements which
fulfill the above conditions.
The above equation 
imposes an order between statements R and S. This direction is 
chosen from their relative order in the program, i.e. if
S appears after R or if R appears after S. 
Finally, the overall independence distance cost function is assembled
from the cost of all pairs $(R,S) \in \mathcal{I}$ with:
\noindent
\begin{equation}
\label{eq:obj:sis}
\min \left\{\nabla_{sum}\right\}
\end{equation}





\subsection{Parallelism on Stencil Computations}
\label{sec:transfos:stencils}

Traditionally, parallelism in time-iterated stencil computations has
been primarily achieved 
through skewing transformations (to enable wavefront parallelization). The approach we take 
is to decompose the stencil optimization into 3 tasks:
i) {\bf Stencil Dependence Classification (SDC)},
ii) {\bf Stencil Parallelism Constraints (SPAR)}
and,
iii) {\bf Stencil Minimization of Vector Skewing (SMVS)}. 
\paragraph{Stencil Dependence Classification (SDC)}
We carefully push dependence satisfaction into specific
scheduling dimensions based on the type of dependences
encountered. The classification applied is quite simple
and stems directly from the inherent structure of the stencil.
The cases that we consider are the following:
i) Non-self forward dependences (NSFD): satisfy at the scheduling 
level corresponding to the time loop (outermost linear dimension);
ii) Non-self backward dependence (NSBD): satisfy at some inner scalar
dimension;
iii) Self-dependences and No. statements $> 1$ (SDN): satisfy
at the second linear schedule dimension (corresponds to the first
spatial dimension)
iv) Self-dependences and No. statements = 1 (SD1): can only be
satisfied by the linear scheduling dimensions, proceed greedily
preferring outer linear dimensions .

\begin{equation*}
{\footnotesize
\boxed{
\begin{split}
MULTI\_SKEW \equiv True: \Psi^{+}_1 = \sum_{\mathcal{D}_{R,S} \in NSFD} \delta^{\mathcal{D_{R,S}}}_1 
\\
MULTI\_SKEW \equiv True, \forall k \in 0..d-1, k~~even:
\\
\Psi^{+}_k = \sum_{\mathcal{D}_{R,S} \in NSBD} \delta^{\mathcal{D_{R,S}}}_k 
~~\wedge~~
\beta^S_k - \beta^R_k - \delta^{\mathcal{D_{R,S}}}_k \geq 0
\\
\forall k \in 0..d-1, k~~ odd, k \neq 3:
\Psi^{+}_{k} = \sum_{\mathcal{D}_{R,S} \in SD1} \delta^{\mathcal{D_{R,S}}}_{k} \\
\Psi^{+}_3 = \sum_{\mathcal{D}_{R,S} \in SDN} \delta^{\mathcal{D_{R,S}}}_3,~~
\sum_{k = 0}^{d} \Psi ^{+}_k = No.deps 
\\
\forall k \in 0..d: \Psi^{-}_k + \Psi^{+}_k = No.deps 
\wedge
0 \leq \Psi^{-}_k, \Psi^{+}_k \leq No.deps
\end{split}
}
}
\end{equation*}

With this simple classification we embed the above 
constraints and objectives, and link the $\delta_k$ variables
to $\Psi_k$ variables. The latter are inserted in the leading position
of the system from the outermost to innermost dimension:
Above we have used the predicate
$MULTI\_SKEW := No.cores < 2 \times OPV$,
$OPV = operations~per~vector$, 
which allows to adjust the degree of skewing w.r.t the
time dimension. The previous constraints allow to
decide {\em where to satisfy dependences}. Next, we will
discuss {\em how} to satisfy them.

\noindent
\paragraph{Stencil Parallelism Objectives (SPAR)}
We now explain how must inter- and intra-dependences be satisfied. 
In the intra- case, we apply a shift along the time dimension. This
boils down to peeling an iteration from the
dependence sources. In addition, it also serves the purpose
of prefetching the data for subsequent iterations.
Next, we enforce a fixed shift along the first
space dimension of the stencil
when the number of cores is ``too large'', and avoid
resorting to skewing the iteration space in this scenario.
The motivation is to avoid large skewing factors (or completely
avoiding them) to minimize the wavefront start-up time required.

\noindent
\begin{equation*}
{\footnotesize
\boxed
{
\begin{split}
\forall \mathcal{D}_{R,S}, R \neq S, ~~IS\_FLOW(\mathcal{D}_{R,S}) \equiv True: 
\beta^S_1 - \beta^R_1 \geq 1 
\\
\forall \mathcal{D}_{R,S}, R \neq S, ~~IS\_FLOW(\mathcal{D}_{R,S}) \equiv True,\\
~MULTI\_SKEW \equiv False: 
\beta^S_3 - \beta^R_3 \geq 2 \times OPV \\
\end{split}
}
}
\end{equation*}

Intra-statement dependences, or self-dependences,
can only be handled by manipulating the coefficients
of the linear scheduling dimensions.
Our approach favors transformations
where the skewing degree reduces level-by-level,
from row-1 to row d-2, and introduce a constraint
to push the skewing factor of the first linear
dimension above a specific threshold (the number
of full-dimensional statements, card(FDS)). This last
constraint is nullified when the statement
in question is not full dimensional. Moreover,
we enforce that the skewing is performed only
w.r.t each space dimension of the stencil:

\noindent
\begin{equation*}
{\footnotesize
\boxed{
\begin{split}
\forall S, ~MULTI\_SKEW \equiv True, ~k \in 0..\lfloor d/2 \rfloor-2, 
\\
min\_dist := \left\{ 1 : k > 0; ~~ 0 : otherwise\right\}:
\\
\left( \sum_{j = 0}^{dim(S)-1}\theta^S_{2k+1,j} \right) -
\left(\sum_{j = 0}^{dim(S)-1}\theta^S_{2(k+1)+1,j}\right)
\geq min\_dist
~~ \wedge
\\
\sum_{j = 0}^{dim(S)-1}\theta^S_{1,j} \geq card(FDS) - BIG\_K . (d - dim(S))
\\
\forall S, ~k \in 1..dim(S)-1:
\theta^S_{2k+1,k} \geq 1 
\end{split}
}
}
\end{equation*}

The next step is to decide if 
skewing is in fact necessary to satisfy the self dependence.
If $\delta^{\mathcal{D}_{S,S}}_3$ is set to 1, then we force
a time skewing on the same dimension, otherwise, 
the time coefficient of the
first spatial dimension can remain as zero. The reader
can observe that if the skewing at level-3 is initiated,
this will force the first linear dimension to have 
a larger skewing factor.
The $\Delta^{-}_{SUM}$ variable is inserted in the leading position
of the current system, while $\Delta^{+}_{SUM}$ is
appended at the end.

\noindent
\begin{equation*}
{\footnotesize
\label{eq:stencils-self-deps}
\boxed{
\begin{split}
\forall S, \mathcal{D}_{S,S}:
\theta^S_{3,0} - \delta^{\mathcal{D}_{S,S}}_3 \geq 0
\\
\Delta^{-}_{SUM} + \Delta^{+}_{SUM} = \sum \delta^{\mathcal{D}_{S,S}}_3
\wedge 
 0 \leq \Delta^{-}_{SUM}, \Delta^{+}_{SUM} \leq No.self.deps
\end{split}
}
}
\end{equation*}

\paragraph{Stencil Minimization of Vector Skewing (SMVS)}
Finally, to avoid the detrimental skewing transformations
that derive in high-stride accesses, we create a cost
function that minimizes that skewing factor of the iterator
and scheduling dimension associated to the fastest-varying
dimension of the dominant array of each statement.
To limit the vector skewing, we use a set of 
$\Phi^S$ variables which are inserted in the leading position 
of the current system.
We call this the "Stencil Minimization for Vector Skewing (SMVS)",

\begin{equation}
{\footnotesize
\label{eq:stencils-vec-skewing}
\begin{split}
\forall S: 
\Phi^S = \sum_{j = 0}^{dim(S)-1} \theta^S_{2(d-1)+1,j} +
\sum_{k = 0}^{d-2} \theta^S_{2k+1,dim(S)-1} 
\end{split}
}
\end{equation}

\subsection{Skewed Parallelism (SKEWPAR)}
\label{sec:skwpar}


A loop is parallel iff it carries no dependence.
To model parallelism, for each statement S and scheduling
dimension $k$,
we use one binary variable $\pi^S_k \in \{0,1\}$.
These variables are associated to all $\delta^{R,S}_k$
and $\delta^{S,R}_k$
dependence satisfaction variables that involve them.
We recall that when a dependence $\mathcal{D}_{R,S}$
is satisfied at some schedule level k, the corresponding
$\delta^{R,S}_k$ is set to 1. In accordance, a loop dimension
surrounding a statement S can be parallelized when 
$\delta^{R,S}_k = 0$ for all possible $\delta$.
We thus upper bound each $\pi^S_k$ in the following manner:

\noindent
\begin{equation*}
{\footnotesize
\label{eq:pivars}
\boxed{
\begin{split}
\forall k \in \{0..d-1\}, \forall \delta^{R,S}_k: 
\pi^R_k \leq 1 - \delta^{R,S}_k \wedge
\pi^S_k \leq 1 - \delta^{R,S}_k 
\end{split}
}
}
\end{equation*}

The benefit of using the introduced $\pi^S_k$ variables is to
centralize the parallelism property and to simplify
several of the proceeding ILP constraints and performance
objectives. Consider for instance, the statement 
$C[i][j] += A[i][k] * B[k][j]$
in the {\em gemm} kernel. Its schedule will have 7 
dimensions. The cannonical loop order for this benchmark,
{\bf (i,j,k)}, will exhibit two outermost parallel loops,
with $\pi_1 = 1 \wedge \pi_3 = 1 \wedge \pi_5 = 0$. 
If however, a transformation
seeks inner parallelism for SIMD-vectorization, then 
$\pi_3 = 0 \wedge \pi_5 = 1$.

It's not always possible to extract outermost parallelism.
The alternative is to focus on extracting this property 
at the second outermost loop dimension,
one surrounded by an outer serial loop
(as in the {\bf cholesky} benchmark).
As a precondition, we require the $\pi$ variables to be previously
inserted. In the event that no other performance objective
has previously created them, we do so here.
To structure the code so as to have parallel loops at the
second linear schedule dimension, we optimize 3 cost functions:
i) maximization of dependence satisfaction at schedule dimension 1;
ii) minimization of sum of coefficients (to minimize potential
skewing transformations induced by (i);
iii) maximization of $\pi$ values at the second linear
dimension (e.q. the second outermost loop).
The combination of these three cost functions allows the scheduler
to choose between loop permutations and skewing transformations
to maximize sync free loops at the second outermost level.
Eq. \ref{eq:skwpar} summarizes these objectives:

\begin{equation}
\label{eq:skwpar}
\left\{ 
\max \sum \delta^{\mathcal{D}_{R,S}}_1, 
\min \sum_{S \in C_i} \sum_{j=0}^{dim(S)-1} \theta^S_{1,j}, 
\max \sum_{S \in C_i} \pi^S_3 \right\}
\end{equation}

\subsection{Space Narrowing (SN)}

In some cases, particularly when we don't expect to be able to 
extract multiple levels of parallelism and to simplify the
ILP solving process, we preset some
scalar schedule coefficients and 
restrict the maximum values of the 
linear coefficients. The former has absolutely no impact on
the correctness of the schedule, whereas the latter
only limits potential skewing transformations.

\noindent
\begin{equation*}
{\footnotesize
\boxed{
\begin{split}
\forall S, N_{SCC} = 1: \beta^S_{d-1} = S \wedge \beta^S_0 = 0
\\
\forall S, N_{SCC} = 1, \forall j \in 0..dim(S)-1: \Theta^S_{1,j} = I^S_{1,j}
\\
\forall S, \forall ~odd~i, \forall j \in 0..dim(S)-1: \Theta^S_{i,j} \leq 2
\end{split}
}
}
\end{equation*}

\subsection{Resource Constrained Optimal Unrolling (RCOU)}
\label{sec:transfos:uaj}
The final step we take is 
a quick analytical exploration to find optimal
unroll and jam factors. This pass is 
applied after the polyhedral code generation process,
i.e. once we recover the classical AST representation.
At this stage, parallel and permutable loops are
marked, and we collect some statement and reference metrics:
\noindent
\begin{equation*}
{\footnotesize
\label{sec:transfos:info}
\boxed{
\begin{split}
resource^S_{j} = \sum_{F \in S} \sum_{i = 0}^{dim(F)-1} |F_{i,j}|
,~
reuse^S_{j} = \sum_{F \in S} |F_{dim(F)-1,j}|,
\\
write^S_j =
\left\{
\begin{matrix}
1 : & j~appears~in~F~\wedge~F~is~write~reference\\
0 : & otherwise
\end{matrix}
\right\}
\end{split}
}
}
\end{equation*}
Then we proceed as follows.
Program $P$ consists of a set of disjoint multi-dimensional
loop nests $\{l_1,l_2,\ldots,l_n\}$. For each loop nest
we first check if its unrollable. The conditions for deeming
a loop nest unrollable are: i) must have at least
one parallel or permutable loop dimension
with constant bounds; ii) $l_i$ can be imperfectly
nested, but statements should only appear in innermost
loops, i.e. $l_i$ can have multiple innermost loops.
If $l_i$ is unrollable we proceed to build the 
{\em resource}, {\em reuse} and {\em write} 
matrices
for all the statements contained in $l_i$.
Next, we produce
a list of nodes contained in $l_i$, and 
set the exploration space of unrolling
factors for each loop dimension
to the set $\{1,2,4,8,16\}$.
If some
loop dimension in $l_i$ doesn't meet the criteria
of condition (i),
then the set becomes the singleton $\{1\}$.
The space of candidate unrolling factors for the entire
loop nest
$l_i$ will be the cartesian
product $UF_1 \times UF_2 \times \ldots UF_n$,
where $n$ is the number of loops in the loop nest.
\begin{scriptsize}
\SetInd{0.4em}{0.5em}
\begin{algorithm}
\KwIn{space: unrolling factors to consider; T: loop nest to unroll; resource, reuse, and write matrices}
\KwOut{optimal unrolling factors for all loops in loop nest T}
opt\_UF = $\{1 \times 1 \times \ldots \times 1\}$;      opt\_reuse = 0\;
\For{each UF $\in$ space}{
  val\_resource = 0;    val\_reuse = 0\;
  \For{each inner loop $L_i \in$ loop\_nest}{
    iters = iterators of loops from $L_i$ to root of T\;
 	\For {each statement S $\in$ $L_i$} {
   	  \For {each array reference F $\in$ S}{
        resource\_F = 1\;
        \For {each it $\in$ iters}{
          \If {({\bf it} appears in F)}{
            resource\_F *= UF(it)\;
          }
          \If {({\bf it} is iterator of inner loop)}{
          	val\_reuse -= UF(it) $\times$ (resource[S][it] - reuse[S][it])\;
          }
          \Else{
             val\_reuse += (MAX\_DEPTH - depth(it) + 1) $\times$ UF(it) $\times$ ( 3 $\times$ reuse[S][it] + write[S][it])\;
          }
          \If {($\exists~~iterator$~~{\bf jt}~~$\in iters$ s.t. {\bf it} $\rightarrow$ {\bf jt}~~or~~{\bf jt} $\rightarrow$ {\bf it})}{
          val\_reuse = $-\infty$\; 
          }
        }
       	val\_resource += resource\_F\;
      } 
    } 
  }
  \If {($\prod_i$ UF(i) $\geq N\_VEC\_REG/2$)}{
 	opt\_reuse = $-\infty$\; 
  }
  \If{(val\_resource $\leq$ N\_VEC\_REG ~{\bf and} val\_reuse > opt\_reuse)}{
 	opt\_UF = UF;    opt\_reuse = val\_reuse\;
  }
}
\Return{opt\_UF}\;
\caption{explore\_space(space,T,resource,reuse,write)}
\label{algo:explore-space}
\end{algorithm}
\end{scriptsize}

Function $explore\_space$ (Algorithm \ref{algo:explore-space})
evaluates each point in {\em space} by
computing both its resource usage (the number of
array references contained in each inner loop of $l_i$,
and the reuse yielded by the candidate unrolling factors.
The {\em resource usage} is product-wise accumulative w.r.t 
the unrolling factor of each loop surrounding
pair \{statement, reference\}. For example, if
reference $A[i][k]$ is surrounded by 3 loops, $i,j,k$,
and this are unrolled by \{2,4,3\}, then the resource
usage increases from 1 to $2 \times 3$. In addition,
our cost model favors unrolling of non-innermost
loops (the innermost loop already has inherent reuse) and
dimensions which impact write-references,
while also penalizing unrolling of loops that induce
high-strides on array references.
Once the optimal set of
unrolling factors has been determined, a new
loop nest $l'$ is produced by unrolling and jamming
every loop in $l$ by $opt\_uf$. The new program
results from concatenating all the unrolled loop nests.
Here we make two notes: i) in our experiments, we set
$N\_VEC\_REG$ to 32;
ii) the purpose of bounding
the product of unroll factors by $N\_VEC\_REG / 2$
is to account for two FMA units on the SKX  processor.

\subsection{Putting it all together}


Not all the constraints presented here are always embedded into the ILP.  The
main conclusion of this work is that when using a single ILP to find a
transformation schedule, resorting to an exhaustive empirical evaluation
should not be needed, but
additional information on the target architecture and program are needed to
select the right objectives and their optimization order. We thus adopt the
classification strategy shown in Eq. \ref{eq:class}, where we use SCoP metrics
to classify the program into one of four types: stencils (STEN), 2-dimensional
kernels (LDLC), dense linear algebra (HPFP), and other. The metrics we use are the
number of dependences in the SCoP ($N_{dep}$), the number of scheduling
dimensions ($dim(\Theta)$) and the number of self-dependences ($N_{self.dep}$).
The function {\em is\_stencil} returns True when at least half of the
statements in the SCoP refer to at least 2 neighboring points in the grid.

\begin{equation}
\scriptsize{
\begin{split}
class(prog) = 
\left\{
\begin{matrix}
STEN :& {\bf if} (is\_stencil(prog) \wedge N_{dep} \leq 3 \times  dim(\Theta)) \\
LDLC   :& {\bf elseif} (dim(\Theta) \leq 5) \\
HPFP  :& {\bf elseif} (N_{SCC} \geq N_{self.dep}) \\
OTHER:&otherwise
\end{matrix}
\right\}
\end{split}
}
\label{eq:class}
\end{equation}

Once we have determined the class of a program, we
select and order the performance objectives depending on the
target architecture.
Table \ref{tab:objectives} lists the performance objectives
embedded into the ILP for each target platform and class of 
benchmarks. When necessary, some objectives are embedded
only when the number of dependences is under a particular
threshold. 

\begin{table}[htb]
\begin{tabular}{|c|c|}
\hline
 Class  & Priority of Performance Objectives ($+\leftarrow\rightarrow-$) \\
\hline
STEN    & SMVS, SDC, SPAR  \\ 
LDLC      & SO, IP, OPIR, SIS, DGF, OP \\
HPFP     & \{SO, IP, OPIR\}($N_{self.dep} \leq N_{SCC}$), SIS, DGF, OP \\
OTHER   & SO ($N_{dep} < 50$), OP, SN \\
\hline
\end{tabular}
\caption{\label{tab:objectives} Cost functions set and priority per 
program class}
\end{table}

The overall selection of performance objectives, and more importantly, their order
of application is mostly determined by a mix of the target 
architecture, specific traits of the program and the ``hardness'' of 
the objective. Table \ref{tab:objdesc} recaps the peformance idioms
described together with the part of the schedule they affect
and the purpose/goal of the idiom.

\begin{table*}[t]
\begin{footnotesize}
\begin{tabular}{|c|c|c|}
\hline
Performance & Schedule & Goal \\
Idiom       & Target   &      \\
\hline
SMVS   & Last linear dimension & Minimize skewing that could affect the vectorization potential \\
SDC &  Two outermost linear dimensions; 2nd scalar dimension & Map dependence satisfaction to specific dimensions \\
SPAR & Two outermost linear dimensions & Produce a linear combination on
outermost linear dimension; \\
& & free the 2nd linear dimension to extract coarse-grained parallelism \\
\hline
SO  & Innermost linear dimension & Minimization of high-strides \\
OPIR & All linear dimensions except the innermost &  Trade parallelism for data reuse\\
\hline
OP & Outermost linear dimension & Attempt to parallelize the outermost loop dimension \\
IP & Innermost linear dimension & Maximize inner parallelism by minimizing $\sum \delta$ \\
\hline
SIS & Outermost scalar dimension & Increase the difference between the values of $\beta^R$ and  $\beta^S$ \\
DGF & All scalar dimensions & Minimize the lexicographic distance between two statements R and S \\
\hline
\end{tabular}
\caption{\label{tab:objdesc} Summary of performance vocabulary}
\end{footnotesize}
\end{table*}

The rule of thumb followed to set the order of objectives is to give
higher priority to more specific performance idioms. In practice,
this translates to placing first the objectives that affect fewer
schedule dimensions. For instace SMVS, SO, and IP only affect the
innermost schedule dimensions, whereas OPIR affects all the linear
dimensions except the innermost. Intuitively, it is easier to
extract outer croase-grained parallelism than innermost/vector-SIMD
parallelism because we have more dimensions to play with. 
The SDC and SPAR constraints and objectives follow the same rationale.
On the other hand, objectives related to locality, SIS and DGF, impact
exclusively the scalar dimensions. Moreover, as $d + 1$ of the $2d + 1$
schedule dimensions are scalar, these set of transformations are ``easier''
than extracting inner parallelism with IP. 

Aside of the above rules of thumb, {\em a priori} knowledge of the
schedule maneuverability is also used to select the objectives. For instance,
iterative stencil computations will consist typically of a single 
{\em Strongly Connected Component} (SCC). As such, attempting to 
separate unrelated statements with the SIS objective is unnecessary.
In a similar fashion, applying the DGF idiom will likey reduce the
potential parallelism or could make it much harder to extract both
coarse and fine-grained parallelism. We note here that in the stencil 
case, the outermost linear dimension of the schedule will represent
a linear combination of the original loop iterators. Thus, this loop
dimension will be inherently serial. Clearly, these objectives
are not needed for the {\em HPFP} and {\em LDLC} program classes; both
of these strongly benefit from the locality objectives (SIS and DGF).
However, the {\em LDLC} program class has only two linear schedule dimensions,
and are very sensitive to fusion/distribution transformations.
In general, fusing a serial and a parallel loop will yield a
serial loop. Although SO and IP only affect the innermost linear dimension,
in most cases there will be fewer transformations that minimize the
number of high-strides than schedules which produce parallel
innermost loops. As an example, consider the {\em DGEMM} benchmark:
exchanging loop-i or loop-j with loop-k will produce an innermost parallel
loop, but only loop-j will minimze the sum of high-strides.
We also point out that some of these transformations 
could serve correcting purposes (e.g. OP in the HPFP and LDLC program classes).
Lastly, the {\em OTHER} program class exists mainly to cover
kernels/benchmarks which are too complex to be handled by the
linear solver used in our experiments, PIP. A better customized
recipe for these benchmarks require more powerful tools such as
GAMS or CPLEX. We thus defer this to later work.

\section{Experimental Evaluation\label{sec:results}}
The entire scheduling approach presented here has been implemented
in a fully automated fashion, in the PoCC compiler \cite{pocc}. 
PoCC is to our knowledge the only
open-source compiler implementing the multidimensional legal scheduling space
we rely on \cite{pouchet.popl.2011}, hence our implementation choice.
We evaluate the effectiveness of our scheduling approach
with the Polybench/C (v3.2) benchmark suite \cite{polybench} on
3 representative processors. 
As baseline for our experiments we use the Pluto compiler
\cite{uday.pldi.2008} (Version 0.11.4-85). 
For each benchmark we generate a number
of variants for each fusion/distribution heuristic
available in Pluto, namely for {\bf maximum fusion (MF)},
{\bf no fusion or maximum distribution (NF)} and
{\bf smart fusion (SF)}. The set of tile sizes 
chosen vary substantially depending on the values of
the problem sizes, and we include powers of 2 and non-powers
of 2. Roughly, the tile sizes chosen for each loop dimension
belong to the set \{1,2,32,50,64,100,128,...\} up to approximately
a fourth part of the problem size, always trying to have at least
one non-power-of-2 tile size for every power-of-2 evaluated.
The final footprint of the tile sizes considered
are such that they all fit in L1 or in L2. We
use the STANDARD Polybench dataset, where the total 
kernel footprint fits in L3.
Experiments were conducted on a 3.3 GHz, 10-core Intel i9-7900X (codename Skylake-X),
with 32 KB L1,  1 MB L2/core, 13.25 MB L3 (shared).
Benchmarks were compiled using GCC 7.2. As a reference,
on this machine the LAPACK DGEMM kernel achieves 0.64 GF/s.



 \begin{table}[t]
 \begin{scriptsize}
 \begin{tabular}{|c|c|c|c|c|c|c|}
\hline
\hline
Benchmark & Our & Pluto &  Pluto  & Pluto   & \%Space & Overall \\
          & GF  & Space &  Best   & Default & Below   & Speedup \\
          &     & Size  &  (GF/s) & (GF/s)  &         &         \\
\hline
\hline
2mm & 185.13 & 2188 & 197.97 & 21.75 & 99.41\% & 0.94\\
3mm & 153.39 & 2188 & 165.03 & 14.13 & 99.59\% & 0.93\\
adi & 5.24 & 568 & 5.63 & 4.29 & 96.48\% & 0.93\\
atax & 4.57 & 769 & 5.81 & 4.61 & 63.07\% & 0.79\\
bicg & 4.57 & 769 & 5.95 & 4.65 & 65.67\% & 0.77\\
cholesky & 6.43 & 1537 & 6.18 & 6.11 & 100.00\% & 1.04\\
correlation & 90.91 & 2188 & 100.39 & 41.43 & 98.58\% & 0.91\\
covariance & 111.11 & 2188 & 102.74 & 41.65 & 100.00\% & 1.08\\
doitgen & 76.70 & 7204 & 75.13 & 29.70 & 100.00\% & 1.02\\
durbin & 0.23 & 769 & 0.28 & 0.23 & 48.11\% & 0.82\\
dynprog & 3.02 & 7204 & 3.58 & 1.38 & 99.99\% & 0.84\\
fdtd-2d & 45.83 & 568 & 18.34 & 6.58 & 100.00\% & 2.50\\
fdtd-apml & 10.07 & 1537 & 10.44 & 6.42 & 97.66\% & 0.96\\
floyd-warshall & 3.54 & 1537 & 12.01 & 9.94 & 43.79\% & 0.30\\
gemm & 201.33 & 2188 & 226.71 & 101.69 & 97.62\% & 0.89\\
gemver & 20.25 & 769 & 11.09 & 7.68 & 100.00\% & 1.83\\
gesummv & 3.20 & 769 & 8.79 & 5.20 & 12.61\% & 0.36\\
gramschmidt & 5.49 & 2188 & 14.18 & 2.49 & 73.49\% & 0.39\\
jacobi-1d-imper & 3.75 & 145 & 11.87 & 2.55 & 53.10\% & 0.32\\
jacobi-2d-imper & 18.52 & 568 & 19.74 & 4.17 & 99.30\% & 0.94\\
lu & 132.56 & 1702 & 71.67 & 45.54 & 100.00\% & 1.85\\
ludcmp & 2.05 & 2188 & 2.03 & 2.01 & 100.00\% & 1.01\\
mvt & 5.33 & 769 & 6.47 & 4.44 & 94.28\% & 0.82\\
reg\_detect & 0.00 & 73 & 2.55 & 1.84 & 0.00\% & 0.00\\
seidel-2d & 12.00 & 568 & 12.80 & 1.15 & 99.82\% & 0.94\\
symm & 55.89 & 1108 & 81.01 & 0.46 & 93.41\% & 0.69\\
syr2k & 54.55 & 2188 & 33.51 & 29.50 & 100.00\% & 1.63\\
syrk & 45.45 & 2188 & 42.06 & 32.51 & 100.00\% & 1.08\\
trisolv & 4.00 & 769 & 8.11 & 7.67 & 59.69\% & 0.49\\
trmm & 2.93 & 730 & 6.74 & 6.05 & 21.10\% & 0.43\\
\hline
\hline
\end{tabular}

 \caption{Polybench evaluation on Intel Skylake (10 cores) with GCC 7.2 \label{tab:skx:gcc}}
 \end{scriptsize}
 \end{table}

The results of our evaluation are shown in 
Table \ref{tab:skx:gcc}.
For each benchmark we report our achieved GF/s, 
the Pluto space size (No.tiled variants $\times$ 3 fusion~heuristics),
the best Pluto GF/s in the space, the default Pluto tile sizes and heuristic
(32x32x...32, -~-parallel and SmartFuse heuristic), 
the percentage of the outperformed
Pluto space (i.e., the fraction of points in the search space that performs slower than the version we produce), and finally
the speedup over the best Pluto variant.
Overall, our transformation schemes
show very strong performance on practically all dense linear algebra (per our classification), 
low-dimensional (LDLC loop nests) and stencil-type of kernels.
The lowest impact that we achieve are in kernels where practically
no parallelism can be extracted without resorting to 
transformations that increase the dimensionality of the loopnest
(i.e {\em trmm, trisolv, ludcmp, floyd-warshall} and {\em cholesky}). Moreover,
the standard dataset of the {\em reg\_detect} kernel
uses a time loop that performs 10000 iterations while using extremely
small space grids. Given these constraints, both our approach
and the Pluto variants were unable to surpass the original
kernel performance.

Regarding stencil computations ({\em adi, fdtd-apml, fdtd-2d, jacobi-1d-imper,
jacobi-2d-imper, seidel-2d}), we observe that although Pluto applies
time tiling
to maximize locality, the requirement of skewing
the iteration domain to extract parallelism has a 
3-fold impact on the transformed code. 
First, uncontrolled skewing induces the creation of complex 
loop bounds that back-end compilers (e.g. GCC or ICC) are unable 
to analyze, producing very low vectorization ratios. Second,
tiling produces much coarser (and fewer) units of work, making it
unprofitable to use multiple threads per core. Third, the parallelism
obtained by skewing transformations takes much longer to be exploited,
as multiple time iterations might be required to reach the steady 
state.

Next, we focus our attention on the lower dimensional benchmarks ({\em atax,
bicg, durbin, gemver, gesummv, mvt} and {\em trisolv}). These kernels have the
property that only a single level of parallelism can be extracted (without
stripmining or tiling). Hence, choosing between inner-vectorizable loops and
outer-parallel loops is usually necessary. Another phenomenon observed in our
experiments is that Pluto's MF heuristic was extremely detrimental to
parallelism, while NF achieved the worst locality.  Hence, Pluto's MF heuristic
offered the best trade-off between these properties.  These benchmarks are also
bandwidth-bound, so tiling them improves the program's locality at the cost of
reducing the available parallelism.  In this benchmark class, we often fall
20\% below the best Pluto variant while outperforming 60-90\% of the
exploration space.

Turning to the dense linear algebra kernels ({\em gemm, 2mm, 3mm, correlation,
covariance, doitgen, gramschmidt, lu, syrk, syr2k, symm}), our approach
consistently outperforms 93\%-100\% of the Pluto variants, and usually
being 5\%-7\% away from the best Pluto performance when we don't 
outperform the whole space.
These benchmarks exhibit several
degrees of data parallelism which is properly exploited by our strategy.  In
particular, we note that trading parallelism for reuse (OPIR objectives) yields
very strong performance on the Skylake processor, effectively making the
computation to boil down to dot-products in all cases except for {\em doitgen}
which presents an outermost parallel loop. Another crucial aspect to achieving
high-performance is our unrolling cost model for data-reuse at the innermost
loop level, parallel dimension to unroll and the potential penalization for
inducing high-strides. 

\begin{figure}[!htb]
\includegraphics[height=1.5in,width=0.48\textwidth]{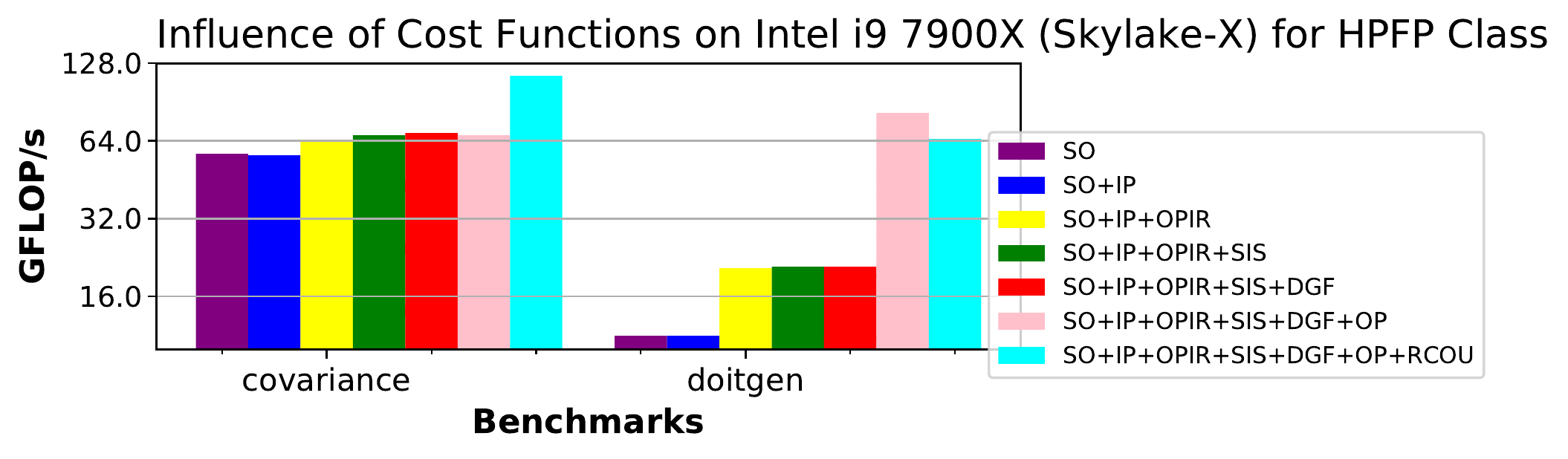}
\includegraphics[height=1.5in,width=0.48\textwidth]{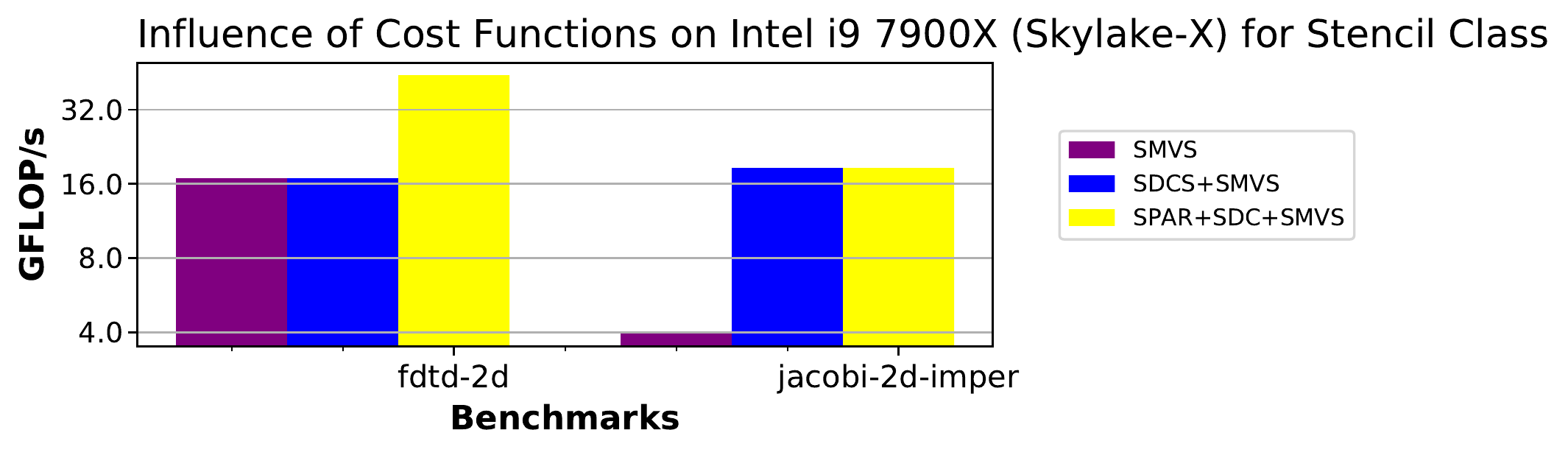}
\caption{\label{fig:levels-skx}Performance influence of transformations on Intel Skylake}
\end{figure}

\paragraph{Performance Influence of Transformations}
To complement our study, we present in Figure \ref{fig:levels-skx}
the cumulative effect of our 
transformations on a subset of benchmarks of the STEN and HPFP classes.
For each benchmark (bar cluster), we show up to 6 levels of 
transformations. 
The exact objectives applied
in each case depends on the architecture and benchmark 
class (See Table \ref{tab:objectives}).
The legend shows objectives added incrementally, from the highest
priority (e.g. SO for {\bf doitgen}) to the lowest one (e.g OP).
The aggregated effect of several performance objectives defined
in our vocabulary are easily observable in the {\bf doitgen}
benchmark, a tensor contraction. In contrast, the same
set of objectives yield marginal performance improvements
for {\bf covariance}. It is interesting to see, however, that 
such a large set of cost functions does not degrade performance
despite seeking seemingly opposing goals. We also observe a slight
performance degradation due to the {\bf RCOU} post-processing,
which practically doubles {\bf covariance}'s performance
while slightly decreasing {\bf doitgen}'s.  For benchmarks
{\bf fdtd-2d} and {\bf jacobi-2d-imper} we see that our
stencil-specific objectives achieve synergistic cooperation
that translate to almost $4\times$ speedup.


\paragraph{Tuning Time}
As last contribution, we show in Table \ref{tab:tuning-time}
the time (sec) to tune a subset of benchmarks
with our framework and with the Pluto compiler.
For our approach, the tuning time consists of the Gen
(transformation), Bin (GCC compilation) and Exec
(kernel execution time for 5 runs). 
For Pluto, we show the average Gen, Bin and Exec times,
which scaled by the space size, gives
the total tuning time. The last column shows the speedup
achieved by our framework w.r.t Pluto's tuning time.

\begin{table}[htb]
{\scriptsize
\begin{tabular}{|c|c|c|c|}
\hline
\multirow{3}{*}{Benchmark} & {\bf Our Approach} & {\bf Pluto Space} & Speed-\\
    & Total Time ( Gen + & Total Tuning Time  & up\\
   & Bin + Exec) & (Gen + Bin + Exec)  & \\
\hline
gemm&3.1 (2+0.29+0.8)&2856.6 (0.16+0.27+0.88) &924.5\\
3mm&88.8 (88+0.55+0.21)&12794.7 (0.55+0.39+4.91) &144.2\\
correlation&675.5 (675+0.37+0.055)&3578.9 (0.77+0.44+0.43) &5.3\\
doitgen&15 (14.45+0.43+0.035)&5806.8 (0.31+0.32+0.18) &389.4\\
fdtd-2d&21.9 (21.2+0.55+0.06)&1220.1 (1.46+0.35+0.34) &56\\
jacobi-2d-imper&7.4 (7.3+0.036+0.027)&1782.3 (0.63+2.43+0.08) &242.1\\
fdtd-apml&1793.7 (1793+0.4+0.3)&3813.1 (1.65+0.39+0.44) &2.2\\
lu&2.6 (2.2+0.27+0.0405)&1600.2 (0.21+0.31+0.42) &637.4\\
gramschmidt&277.6 (277+0.32+0.2445)&2626.8 (0.37+0.26+0.57) &9.5\\
symm&8.1 (7.3+0.51+0.24)&60962.4 (9.11+44.4+1.51) &7573\\
gemver&2.7 (2.3+0.27+0.0395)&819.6 (0.16+0.25+0.66) &314.1\\
\hline
\end{tabular}
}
\caption{\label{tab:tuning-time}Tuning time (sec) for selected benchmarks using GCC 7.2}
\end{table}

\section{Related Work}
This work builds largely on previous research that constructs the convex space
of semantics preserving transformations
\cite{vasilache.07.phd,pouchet.popl.2011}, also known as the Single-ILP
scheduling approach.  In contrast to the level-by-level scheduling strategy
used in compilers such as Pluto \cite{uday.pldi.2008} and PPCG
\cite{ppcg.taco.2013}, leveraging the function space created by applying
the affine form of Farkas lemma, allows to select transformation schedules
that exhibit strongly desired program properties 
that impact performance. Older seminal works such as Feautrier's
\cite{feautrier.ijpp.1992b} minimal-delay schedules 
greedily satisfy dependences at the outermost loop dimensions,
favoring transformations that exhibit inner parallelism. 
Trifunovic's et al. \cite{trifunovic.pact.2009} proposed 
transformations for vectorization, but that only accounted
for loop permutation. Kong et al. \cite{kong.pldi.2013}
developed a complex transformation framework with the
goal of exploiting multi-dimensional reuse of point loops
in tiled programs by integrating permutability,
inner parallelism and stride-0/1 constraints into a single ILP.
However, unlike us, they relied on a highly tuned
and architecture specific SIMD code generator.
Vasilache et al. \cite{vasilache.impact.2012} proposed
a scheduling strategy to simultaneously model
inner parallelism, contiguity and the data layout for
vectorization. Similar to our
work, it first built the convex affine space of all
legal schedules \cite{vasilache.07.phd}. Their formulation
favors stride patterns amenable to SIMD-vectorization 
(stride 1 or 0) while aggregating costs of all array
references.

Only a handful of works resemble our
dependence-oriented objectives. 
Zinenko's et al. \cite{zinenko.tr.2017,zinenko.cc.2018} 
introduced ``spatial proximity'' relations
in a unified template to model contiguity,
cache line access and avoid false sharing. Previously, Bondhugula's
work \cite{uday.sc.2012}
exploited the uniformity of dependences 
in stencil computations to produce ``diamond'' shaped
tiles that maximized its concurrent start and execution.

CHiLL \cite{chen.cgo.2005,hall.lcpc.2009,tiwari.ipdps.2009} is a polyhedral
autotuning framework designed for extensive experimental exploration
via the specification of transformation recipes
(including parameters such
as tile sizes, unrolling factors, among others).
Other
approaches combine (experimental) iterative and model driven 
schemes. Pouchet et al. \cite{pouchet.sc.2010}, explored
fusion heuristics and statement interleavings that lead to substantial
performance improvements.  In a similar line, 
\cite{pouchet.popl.2011} proposed a technique that operated on the legal space
to build all the possible tileable interleavings of statements. 

Domain specific autotuning has also been a subject of research.
Works such as \cite{kamil.ipdps.2010,patus.ipdps.2011,stock.pldi.2014} implemented autotuning frameworks
for stencil computations targeting accelerators and multi-core processors.
Their auto-tuning space included domain decomposition for GPUs, register
blocking, NUMA-aware memory allocation.
Shirako et al. \cite{shirako.sc.2014} 
combined polyhedral and AST transformations by using a cache line cost model,
while
Vasilache et al. \cite{vasilache.ics.2005} designed a mechanism
for decomposing, isolating and recomposing the transformation effects of a
schedule.

\label{sec:related}

\section{Conclusion}
Traditional approaches for loop transformations via polyhedral scheduling
attempt to find the best one-size-fits-all set of objectives to achieve good
overall performance, yet the state of practice is to employ auto-tuning to
ensure the best performance on different architectures.  In this work we take a
significantly different approach, by instead creating a common set of
performance objectives, or idioms, which serve as bases for a vocabulary.
Using this high-level vocabulary, we craft transformation recipes
by selecting and embedding specific idioms into  in a single-ILP.
The objectives selected are made so to exploit both program characteristics
and architectural target traits. 
Although building the transformation recipes  entail a mix of
performance engineering and a bit of empirical evaluation, we are
still able to avoid, to a large extent, large experimental evaluation.
We also provide several high-level intuitions and rationales that
justify both the selection of performance idioms and their order
of application.

Another result of our work is how simple SCoP metrics (e.g. No. of SCC or No.
of self dependences) can be used to easily categorize programs.  This
classification, in tandem with some prior knowledge of the target architecture
allows us to wisely select the adequate objectives to embed in the ILP, and
more importantly, to prioritize them. 
We proposed
novel objectives such as the
trade-off between {\em Outer Parallelism - Inner 
Reuse} (OPIR), {\em Stride Optimization} (SO),
{\em Separation of Independent Statements} (SIS),
{\em Dependence Guided Fusion} (DGF), 
together with constraints specific to stencils 
computations that limit skewing,
select the dimension at which a dependence
is satisfied and how to satisfy it. 
Our experimental evaluation shows that strong performance can be achieved
at almost no tuning cost by using the performance vocabulary presented here.

\label{sec:conclusion}

\bibliographystyle{ACM-Reference-Format}
\bibliography{martin.bib}

\end{document}